\documentclass[%
 reprint,
 amsmath,amssymb,
 aps,
]{revtex4-2}

\usepackage{graphicx} 
\usepackage{ulem}
\usepackage{float}
\usepackage{hyperref}

\usepackage{xcolor}
\usepackage{amssymb}
\usepackage{dcolumn}

\begin{document}
\preprint{APS/123-QED}

\title{
Electrostatic control of quantum phases in KTaO$_3$-based planar constrictions
}

\author{Jordan T. McCourt}
\author{Ethan G. Arnault}
\author{Merve Baksi}
\author{Divine P. Kumah}
\author{Gleb Finkelstein}
\thanks{Corresponding author. Email: gleb@duke.edu}
\affiliation{%
 Department of Physics, Duke University, Durham, NC 27708, USA
}%

\author{Samuel J. Poage}
\author{Salva Salmani-Rezaie}
\author{Kaveh Ahadi}
\affiliation{%
 Department of Materials Science and Engineering, The Ohio State University, Columbus, Ohio 43210 USA
}%

\date{\today}

\begin{abstract}
Two-dimensional electron gases (2DEGs) formed at complex oxide interfaces offer a unique platform to engineer quantum nanostructures. However, scalable fabrication of locally addressable devices in these materials remains challenging. Here, we demonstrate an efficient fabrication approach by patterning narrow constrictions in a superconducting KTaO$_3$-based heterostructure. The constrictions are individually tunable via the coplanar side gates formed within the same 2DEG plane. Our technique leverages the high dielectric permittivity of KTaO$_3$ ($\epsilon_r\sim5000$) to achieve strong electrostatic modulation of the superconducting 2DEG. Transport measurements through the constriction reveal a range of transport regimes: Within the superconducting state, we demonstrate efficient modulation of the critical current and Berezinskii–Kosterlitz–Thouless (BKT) transition temperature at the weak link. Further tuning of the gate voltage reveals an unexpectedly regular Coulomb blockade pattern. All of these states are achievable with a side gate voltage $|V_\mathrm{SG}|\lesssim 1$ V. The fabrication process is scalable and versatile, enabling a platform both to make superconducting field-effect transistors and to study a wide array of physical phenomena present at complex oxide interfaces.
\end{abstract}

\maketitle

\section{Introduction}

\begin{figure*}[htp!]
\centering
\includegraphics[width=\textwidth]{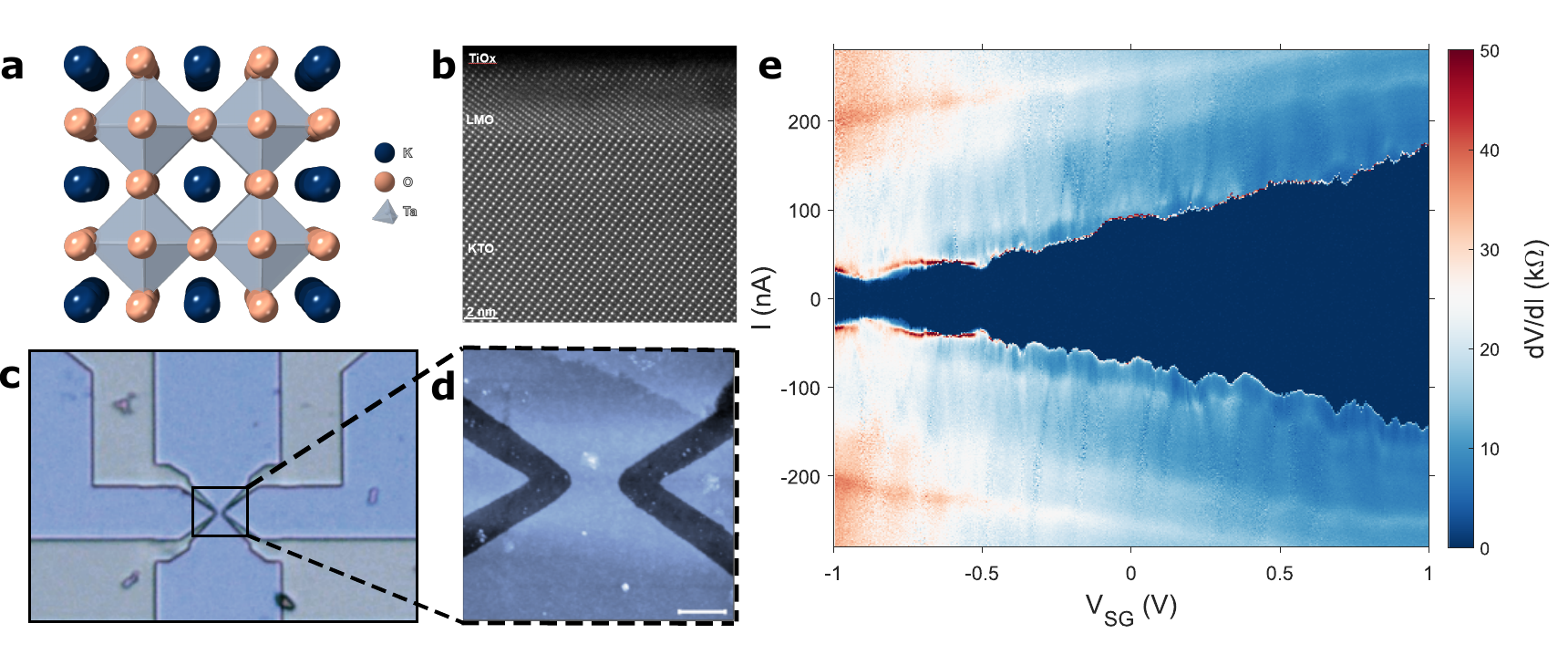}
\caption{\textbf{(a)} Lattice structure of the cubic perovskite KTaO$_3$. \textbf{(b)} A HAADF-STEM image of the sample with LaMnO$_{3}$ grown on a KTaO$_{3}$ (111) substrate and capped with TiO$_{x}$. The highly crystalline structure of the KTaO$_{3}$ and LaMnO$_{3}$ can be clearly seen. \textbf{(c)} Optical image of a KTaO$_{3}$ heterostructure with the developed lithographic mask, prior to etching the device. \textbf{(d)} An AFM image of the weak link measured in Figures \ref{fig:1}e, \ref{fig:IcRn} and \ref{fig:temperature}. The constriction is formed by two insulating trenches (dark regions in the image) which separate it from the coplanar side gates formed in the remainder of the film. Scale bar is 1$\mu$m. \textbf{(e)}  Map of the differential resistance measured across the constriction as a function of $I$ and $V_\mathrm{SG}$ in the $n\approx 5.7\times10^{13}$ cm$^{-2}$ device at 120 mK.}
\label{fig:1}
\end{figure*}

Two-dimensional electron gases (2DEGs) formed at the interfaces of complex oxides have emerged as a promising platform to explore the interplay of superconductivity, spin-orbit coupling, nematicity and ferromagnetism~\cite{hwang_emergent_2012}. The recent discovery of  superconductivity at the (111) interfaces of KTaO$_3$ is particularly interesting due to its relatively high critical temperature~\cite{liu_two-dimensional_2021, chen_electric_2021, qiao_gate_2021}, anisotropic in-plane superconductivity~\cite{liu_two-dimensional_2021,arnault_anisotropic_2023,zhang_spontaneous_2023}, giant critical magnetic fields~\cite{al-tawhid_enhanced_2023, filippozzi_high-field_2024}, low superfluid stiffness~\cite{mallik_superfluid_2022,yang_tuning_2025,yu_sketched_2025} and ferromagnetic order~\cite{krantz_intrinsic_2024,hua_superconducting_2024,ning_coexistence_2024}.

Critically, these phenomena display a strong dependence on the carrier density. Fortunately, the KTaO$_3$ 2DEGs are highly tunable because of their relatively low carrier density (several times $10^{13}$ cm$^{-2}$), and the high dielectric permittivity of the substrate $\sim 5000$)~\cite{geyer_microwave_2005}. This combination of properties allows one to tune the electron density over a wide range between the superconducting and insulating states~\cite{chen_electric_2021}. Similar tunability is also present in 2D van der Waals crystals and Moir\'e superlattices, however such structures require complex fabrication techniques that are hard to control, and often result in significant device-to-device variations~\cite{lau_reproducibility_2022}. Overall, the combination of electrostatic field control and scalable growth make KTaO$_3$ a highly promising platform for achieving fully integrated superconducting electronics~\cite{mannhart_oxide_2010,schlom_perspective_2015,coll_towards_2019,kim_electronic-grade_2024}. 

Despite this promise, so far the electrostatic modulation of the superconducting state in KTaO$_3$ has only been achieved by applying back gate voltage of up to 200 V to `switch off' superconductivity~\cite{chen_electric_2021}. The high voltage requirement greatly limits the practical application of this promising material. Here, we overcome the limitation by presenting a novel fabrication route for making KTaO$_3$ devices, which display a wide range of transport properties with a gate voltage less than 1 V in magnitude. Specifically, we etch narrow constrictions in the plane of the structure, which serve as superconducting weak links. We follow the classical Dayem bridge design of Ref.~\cite{anderson_radio-frequency_1964} with a notable modification: the constrictions are defined by narrow trenches, and the disconnected regions of the film serve as self-aligned side gates.
\\
\indent The gates are highly efficient in tuning the junctions, allowing us to explore several transport regimes within one structure. We present transport measurements which demonstrate the electrostatic tuning of the superconducting state within the constriction via large modulation of the critical current. Furthermore, the constrictions demonstrate the Berezinskii–Kosterlitz–Thouless transition with gate-tunable critical temperature. Further depletion of the constrictions results in a dissipative regime in which the resistance can be varied by 4 orders of magnitude. There, we can resolve a regular pattern of conductance peaks corresponding to the Coulomb blockade of Cooper pairs. This evolution establishes KTaO$_3$ as a platform for studying the interplay of multiple quantum phenomena, all in a single device. The results presented here could also be relevant to the field of quantum devices, as we present a simple, scalable method to fabricate all-superconducting field effect transistors.

\section{Results}

The heterostructures are grown via the oxide molecular beam epitaxy, starting with a KTaO$_3$ (111) substrate, followed by a layer of LaMnO$_3$, and capped by TiO$_x$~\cite{al-tawhid_enhanced_2023}. A 2-dimensional electron gas (2DEG) is formed at the interface of LaMnO$_3$ and KTaO$_3$ (Figure~\ref{fig:1}c). The capping layer of TiO$_x$ serves to modify the initial carrier density via oxygen depletion. The resulting carrier density, $n$, is in the mid-$10^{13}$ cm$^{-2}$ range, which is high compared to the typical 2DEGs in semiconducting heterostructures but very low compared to metal films.

Our devices are defined by a single round of electron-beam lithography and wet etching. After the development of the e-beam resist (PMMA), we perform a selective wet etch with a KI/HCl mixture to remove the exposed layers grown on top of the KTaO$_3$ substrate~\cite{bridoux_alternative_2012}. Simultaneously, the etch oxidizes the underlying surface, leaving it insulating, while the regions protected by the e-beam resist remain conductive. With this technique, we pattern narrow ($\sim$ 1 $\mu$m wide) constrictions separated from the rest of the electron gas by insulating trenches (also $\sim$ 1 $\mu$m wide). 

The conductive regions flanking the constrictions are used as the side gates (Figure~\ref{fig:1}d). A very similar design has been pioneered in a LaAlO$_3$/SrTiO$_3$ heterostructure~\cite{monteiro_side_2017}. Unfortunately, the tunability of those constriction was limited, whilst more tunable devices in that material system required complex fabrication techniques, e.g.~\cite{thierschmann_transport_2018}. In our case, strong modulation is achieved by side gate voltages $|V_\mathrm{SG}|\lesssim 1$ V. Assuming an approximate field-induced density variation of $\delta n \sim \frac{\epsilon_r\epsilon_0}{ed}V_\mathrm{SG} $ (where $e$ is the electron charge and $d$ is the distance between the gate and the channel), we estimate a tunability of a few $\times10^{13}$ cm$^{-2}$V$^{-1}$ in our devices. Note that these large variations of density are enabled by the very high dielectric constant of KTaO$_3$, which approaches 5000~\cite{geyer_microwave_2005}. The dielectric constant falls with electric field, however it should still stay in the range of a few thousand for the largest fields we apply in this work (on the order of 10 kV/cm)~\cite{wemple_transport_1965,fujii_dielectric_1976}.

We present transport measurements of two sets of constrictions made in the same way and on the same substrate. Figures 1-3 show one of the constrictions from the first set. Here, the bulk electron density was $n\approx 5.7\times10^{13}$ cm$^{-2}$.
Following the measurements, the sample was left in ambient conditions before fabricating the second set of constrictions (Figure 4), which resulted in density reduction to $n \approx 3.6\times10^{13}$ cm$^{-2}$. The bulk critical temperature of the as-grown heterostructure was $T_C\approx 1.2$ K, indicating a BCS gap of $\Delta_\mathrm{BCS} \approx 1.76 k_B T_C \approx 180 \mu$eV. The materials used for the second set of constrictions had a significantly suppressed bulk $T_C \approx 0.45$ K, as expected for the reduced density~\cite{liu_tunable_2023}. 

We begin by probing the transport properties of the first device, measured at 120 mK in a four-probe configuration. In Figure~\ref{fig:1}d, we display the complete map of $\frac{dV}{dI}$ as a function of bias current $I$ and $V_\mathrm{SG}$. The central dark region corresponds to the supercurrent. The measurement displays a largely monotonic increase of $I_C$ with increasing $V_\mathrm{SG}$. In contrast to work in similar systems~\cite{mikheev_quantized_2021}, we do not observe signatures of quantization in $I_C$ ($\delta I_C = \frac{e\Delta}{\hbar}$) despite approaching the limit of $\delta I_C$. We attribute the lack of quantized steps in $I_C$ to the fact that the mean free path is much smaller than the width of the constriction.

\begin{figure}[t]
\centering
\includegraphics[width=0.5\textwidth]{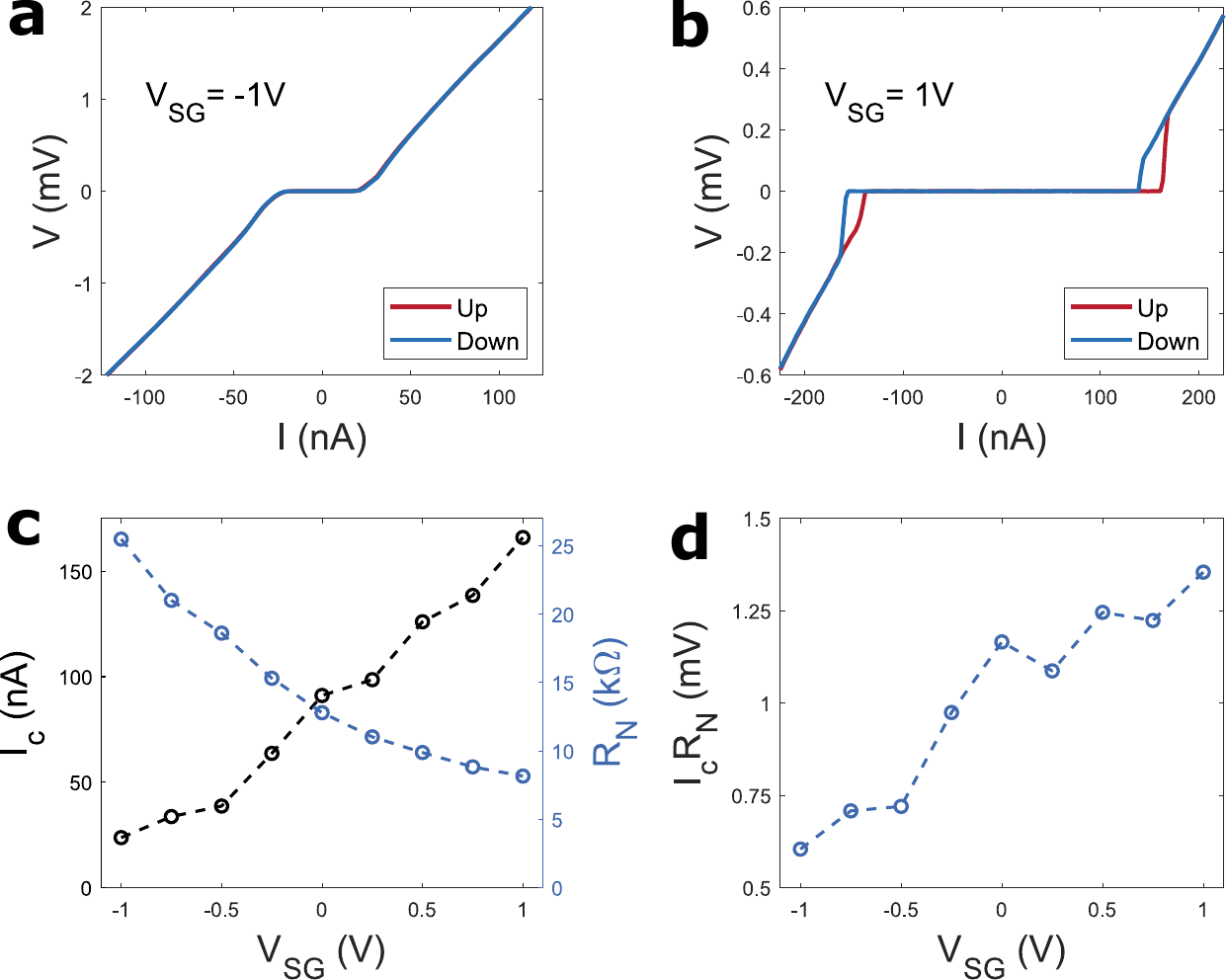}
\caption{\textbf{(a,b)} I-V traces measured at $V_\mathrm{SG}=-1$ V and $1$ V respectively in the $n\approx 5.7\times10^{13}$ cm$^{-2}$ device at 120 mK. \textbf{(c)}~Extracted values of $I_C$ and $R_N$ as a function of $V_\mathrm{SG}$. Induced by the electrostatically modified $n$, the two quantities vary roughly inversely to each other. \textbf{(d)}~$I_CR_N$ calculated from panel \textbf{c}. The product greatly exceeds $\Delta_\mathrm{BCS}$ (estimated to be $\approx 180 \mu$eV), which is  attributed to the geometrical properties of the constriction.}
\label{fig:IcRn}
\end{figure}

Changing $V_\mathrm{SG}$ from $-1$ V to 1 V, we observe a dramatic increase of the critical current ($I_C$) at the constriction, from  a minimum of 15 nA to 170 nA (see Figure~\ref{fig:IcRn}a \& b for IV curves). Concurrently with the change in $I_C$, the normal state resistance, $R_N$, decreases from $\approx25$ k$\Omega$ at $V_\mathrm{SG}=-1$ V to $\approx7$ k$\Omega$ at $V_\mathrm{SG}=1$V (Figure~\ref{fig:IcRn}c). We attribute the growth of the critical current and the decrease of resistance to the increase of $n$ at the constriction via the electrostatic field effect. We find that $I_C$ and $R_N$ vary roughly inversely to each other with changes in $V_\mathrm{SG}$. Their product, $I_C R_N$, is found to increase with $V_\mathrm{SG}$ from a minimum of 0.6 mV to a maximum of 1.35mV (Figure~\ref{fig:IcRn}d). We note that measured values of $I_CR_N$ reach up to $\approx7.5 \Delta_\mathrm{BCS}$, significantly larger than the value of $\approx2\Delta_\mathrm{BCS}$  predicted by Kulik-Omel'yanchuk theory in the dirty limit~\cite{tinkham_introduction_1996}. While unconventional superconductivity may contribute to the enhancement of the $I_C R_N$ product, the most likely explanation is that the $I_CR_N$ product reflects the geometric properties of the device given that the constriction length is much larger than the coherence length ($\sim10$ nm).

\begin{figure}[b]
\centering
\includegraphics[width=0.5\textwidth]{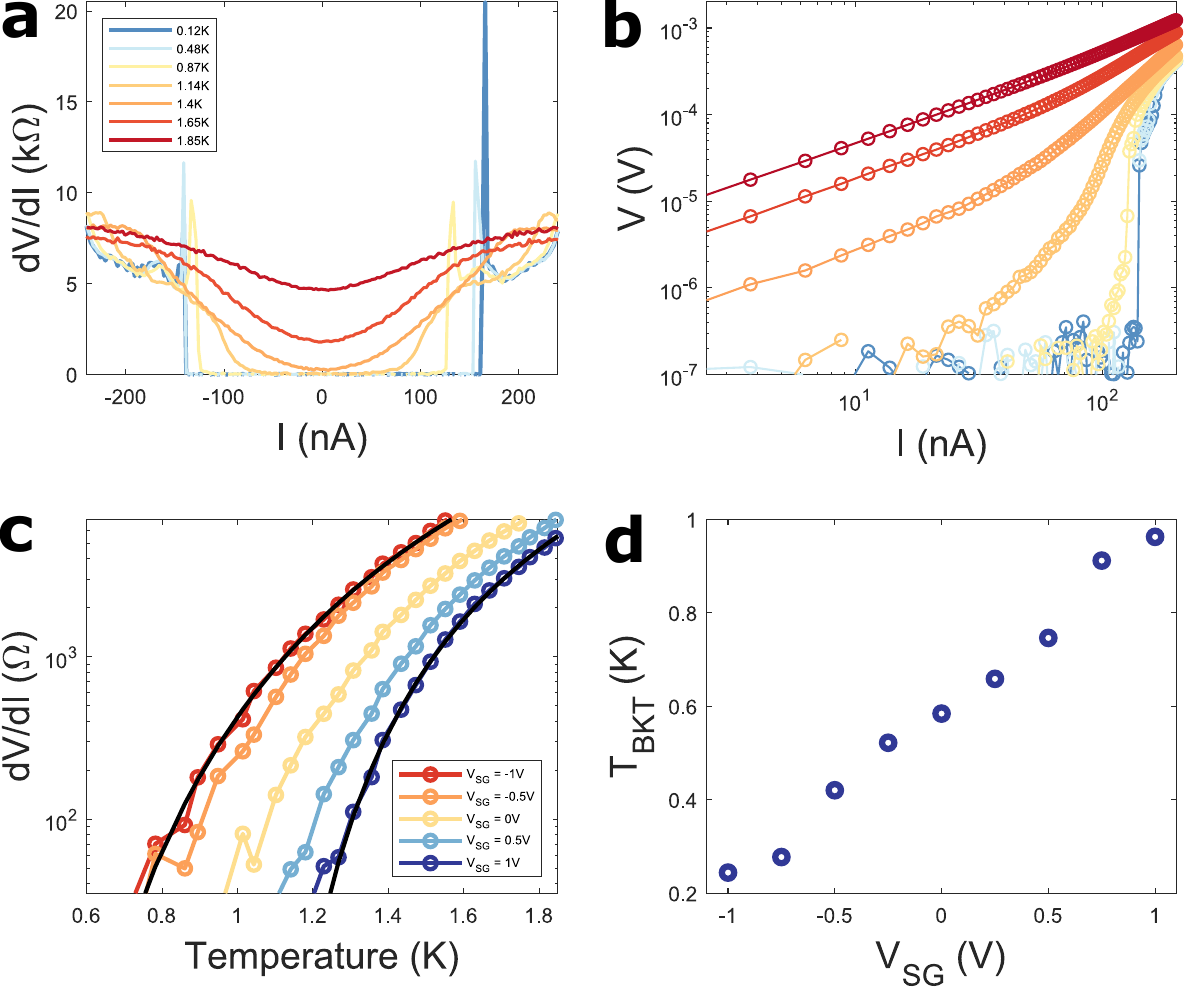}
\caption{\textbf{(a)} $R=\frac{dV}{dI}$ measured at $V_\mathrm{SG}=1$ V and several temperatures in the $n\approx 5.7\times10^{13}$ cm$^{-2}$ device. \textbf{(b)} The corresponding I-V curves, measured at the same temperatures and the same  $V_\mathrm{SG}=1$ V. A transition from high powers of $I$ to the linear regime is observed as temperature is increased, suggesting that the $T=1.14$ K curve is close to the transition. \textbf{(c)} Dependence of the $R(T)$ curves on the gate voltage. Fits to the Halperin-Nelson formula are overlaid on the curves measured at $V_\mathrm{SG}=-1$ and 1 V. \textbf{(d)} Extracted values of $T_\mathrm{BKT}$ plotted vs  $V_\mathrm{SG}$.}
\label{fig:temperature}
\end{figure}

We now switch to the dependence of the transport through the constriction on temperature. Figure~\ref{fig:temperature}a shows differential resistance $R=\frac{dV}{dI}$ measured vs $I$ at temperatures between 0.12 K and 1.85 K. 

The width of the constriction is significantly (several tens to a hundred times) larger than the coherence length, so that we could approximately consider it as a 2D film. The Berezinskii-Kosterlitz-Thouless (BKT) transition could be expected in this case, which has been previously studied in bulk KTO$_3$ films in Ref.~\cite{liu_tunable_2023}.
The temperature dependence of the I-V curves allows us to roughly identify the transition temperature, whereby $V\propto I^3$ (Figure~\ref{fig:temperature}b). $T_\mathrm{BKT}$ can be more precisely extracted from the $R(T)$ curves by fitting them with the Halperin-Nelson (HN) formula~\cite{halperin_resistive_1979}. Figure~\ref{fig:temperature}c shows $R(T)$ for several $V_\mathrm{SG}$. We fit each curve with the HN formula, shown in black solid lines for the two extreme cases in panel (c). Despite the fact that the density varies across the constrictions, the fits work very well, which allows us to extract the dependence of $T_\mathrm{BKT}$ on $V_\mathrm{SG}$, Figure~\ref{fig:temperature}d.  $T_\mathrm{BKT}$ is found to vary approximately linearly from the maximum of $\approx1$ K to the minimum of $\approx 0.2$ K as we tune $n$. The measurement highlights the ability of our devices to controllably probe the physical properties of the 2DEG in an easily accessible gate voltage range.
 
We note that it has been suggested that the presence of inhomogeneous superconducting islands in oxide heterostructures may explain the non-linear I-V characteristics \cite{venditti_nonlinear_2019} otherwise attributed to BKT physics. However, this explanation seems less likely in the constriction studied in Figure~\ref{fig:temperature}, which demonstrates a well defined and monotonic supercurrent across most of the gate voltage range shown in Figure~\ref{fig:1}e. However, formation of a superconducting island was detected for the film which has a lower electron density, as we describe now. 

\begin{figure}[!htbp]
\centering
\includegraphics[width=0.5\textwidth]{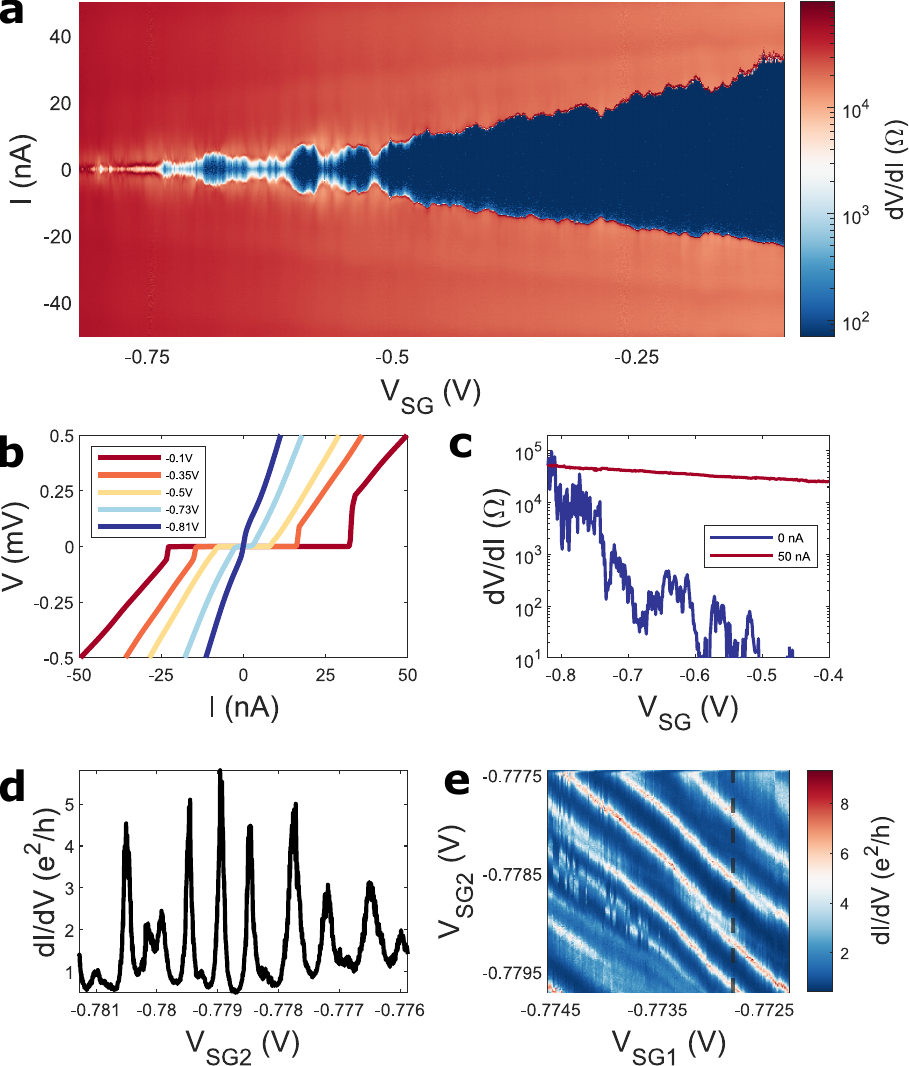}
\caption{\textbf{(a)} Map of $R=\frac{dV}{dI}$ across the $n=3.6\times10$ cm$^{-2}$ device measured at 65 mK as a function of $I$ and $V_\mathrm{SG}$. Similar to Figure~\ref{fig:1}e, $I_C$ decreases with decreasing $V_\mathrm{SG}$, but here it approaches zero $V_\mathrm{SG}\lesssim-0.5$ V
and the constriction enters a fully dissipative regime. \textbf{(b)} I-V curves corresponding to cuts of panel \textbf{a} at several $V_\mathrm{SG}$. The supercurrent is gradually suppressed until the last curve, which displays the presence of an insulating state with zero-bias resistance exceeding the normal resistance. \textbf{(c)} $R(V_\mathrm{SG})$ measured at $I=0$ and 50 nA. For $I=0$, $R$ decreases with increasing $V_\mathrm{SG}$ until $\sim-0.5$ V, at which point the noise floor is reached in the superconducting regime. \textbf{(d)} Zero-bias conductance $G=\frac{dI}{dV}$ as a function of $V_{SG2}$ whilst $V_{SG1}$ is kept constant at $-0.7728$ V (the value is marked the dashed line in panel \textbf{e}). Nearly periodic resonant conductance peaks are observed. \textbf{(e)} Zero-bias conductance map measured vs $V_{SG1}$ and $V_{SG2}$. The slope of the resonant features is close to 1:1, indicating a similar coupling of the dot to the two gates. } 
\label{fig:junction2}
\end{figure}

Figure~\ref{fig:junction2}a presents the differential resistance map of the constriction made in the film with the reduced density of $n=3.6\times10$ cm$^{-2}$. Although the device geometry is the same as in Figure 1, now its normal resistance could be tuned in excess of several tens of kOhm (Figure~\ref{fig:junction2}c). Simultaneously, the supercurrent branch in the I-V curves is suppressed, Figure~\ref{fig:junction2}b. 
Eventually, we observe I-V curves with zero-bias resistance exceeding normal resistance, which suggest a transition to an insulating state (see the curve at $V_\mathrm{SG}=-0.81$ in Figure~\ref{fig:junction2}b). The gate-tunable transition between the superconducting and the insulating states in an all-superconducting constriction highlights the potential of using KTaO$_3$-based devices in future applications.

We have explored the dissipative regime by measuring the zero-bias differential conductance $G=\frac{dI}{dV}$ by voltage biasing the sample, see Figure~\ref{fig:junction2}d. There, we observe resonant features which we attribute to a Coulomb blockade effect as Cooper pairs tunnel onto a superconducting island~\cite{tinkham_introduction_1996}. The peaks' position can be individually tuned by each side gate, (Figure~\ref{fig:junction2}e). Interestingly, the features follow a nearly diagonal slope, which demonstrates an approximately equal coupling to the two side gates, indicating that the resonances originate in the middle of the constriction.  This, and the relative regularity of the pattern are unexpected and suggest that the resonances are not formed due to localization of the carriers in multiple disorder-induced puddles. While the exact orgin of the resonances is difficult to establish, we note that the highly non-linear dielectric constant of KTaO$_3$ may favor formation of a conducting channel in the middle of the constriction.

\section{Discussion}
Our results demonstrate an efficient method to fabricate highly tunable KTaO$_3$-based superconducting transistors. While we follow multiple prior demonstrations of quantum devices in oxide-based heterostructures~\cite{ron_anomalous_2014, gallagher_gate-tunable_2014, bal_gate-tunable_2015, goswami_quantum_2016, stornaiuolo_signatures_2017,thierschmann_transport_2018, mikheev_quantized_2021, singh_gate-tunable_2022, yu_nanoscale_2022,  yu_sketched_2025, cen_oxide_2009}, most of these pioneering works required complex, multi-step fabrication techniques. Here, the constrictions were made simply via electron beam lithography followed by a wet chemical etch. Based on the feature size ($\approx1\mu$m), such devices should be compatible with optical lithography, greatly increasing scalability and accessibility. The coplanar geometry presented here may also be less likely to suffer from bulk-induced disorder and complex trapping mechanisms, as has been observed in back-gated devices~\cite{yin_electron_2020}.

The detection of the periodic conductance peaks and an insulating state at negative $V_\mathrm{SG}$ demonstrates the versatility of the side gates in accessing multiple transport regimes. It is very likely that further narrowing of the constriction would enable a transition to a resistive state even in devices with a higher $T_C$. As a result, geometries fabricated with the technique presented here may be suitable to study field-induced transitions and the interplay of superconductivity and spin-orbit coupling in low-dimensional systems.

\section{Summary}
In summary, we have developed a simple, scalable process to fabricate micron-scale superconducting Dayem bridges at the KTaO$_3$ interface. The coplanar side gates allow us to  tune the carrier density at the constriction within an easily accessible gate voltage range. We first demonstrate an efficient control of the BKT transition over a wide range of temperatures. By further quenching the carrier density, we can pass into a dissipative regime, before ultimately reaching a highly resistive state with resonant conductance peaks. The ability to tune a single device between these regimes is unprecedented in oxide electronics. It highlights the promise of KTaO$_3$ as an exciting playground for studying numerous exotic phenomena in a single, highly tunable device.

\section{Notes}
\noindent Whilst preparing this manuscript, we became aware of a work also using etch-defined side gates to control electron density in a KTaO$_3$-based heterostructure~\cite{kimbell_electric_2025}.

\section{Materials and Methods}
The devices were fabricated in two steps. We first used electron beam lithography (EBL) to define bonding pads of 5/50nm of Cr/Au, which was deposited by thermal evaporation at a base pressure of $2\times10^{-7}$ Torr. The device with a conductive polymer to enable the lithography. 
The second step uses EBL to define the mesa and the constrictions, which are then wet etched with a HCl/KI mixture. We stress that the device can be fabricated with just the second lithography step, however wire-bonding to the device proved challenging without visible bonding pads.

The transport measurements were carried out in a Leiden Cryogenics dilution refrigerator. The measurement lines were fitted with 2nd-order low-pass filters, thermalized to the mixing chamber plate, to avoid spurious microwave frequency radiation from heating the sample. Most measurement were performed in a four-probe configuration, except for the second half of Figure~\ref{fig:junction2}. In the latter case, the filter and line resistance of 10.01 kOhm was subtracted from the two-probe differential conductance measurements.
Differential resistance was measured with a 0.25-3 nA excitation
and the differential conductance was measured with a 0.5 $\mu$V excitation using homemade electronics. 

\begin{acknowledgments}
Transport measurements by J.T.M. and data analysis by J.T.M., E.G.A. and G.F. were supported by the National Science Foundation grant DMR-2327535. K.A and S.J.P were supported by the U.S. National Science Foundation under Grant No. NSF DMR-2408890. Film synthesis by D.K. and M.B. were supported by the U.S. National Science Foundation under Grant No. NSF DMR-2324174. Sample fabrication by J.T.M. was performed in part at the Duke University Shared Materials Instrumentation Facility (SMIF), a member of the North Carolina Research Triangle Nanotechnology Network (RTNN), which is supported by the National Science Foundation (award number ECCS-2025064) as part of the National Nanotechnology Coordinated Infrastructure (NNCI).

\end{acknowledgments}

\bibliographystyle{unsrt} 
\bibliography{references} 

\end{document}